\documentclass{jpsj3}

\usepackage{amsmath}

\def\Vec#1{\mbox{\boldmath $#1$}}
\makeatletter
\newcommand{\LEQQ}{\mathrel{\mathpalette\gl@align<}}
\newcommand{\GEQQ}{\mathrel{\mathpalette\gl@align>}}
\newcommand{\gl@align}[2]{\lower.6ex\vbox{\baselineskip\z@skip\lineskip\z@\ialign{$\m@th#1\hfil##\hfil$\crcr#2\crcr=\crcr}}}
\makeatother

\title{Nonequilibrium Molecular Dynamics Simulation of Interacting Many Electrons Scattered by Lattice Vibrations}

\author{Fan LEE\thanks{E-mail address: lee@ASone.c.u-tokyo.ac.jp} , 
        Tatsuro YUGE$^1$\thanks{E-mail address: yuge@m.tohoku.ac.jp}  
        and Akira SHIMIZU\thanks{E-mail address: shmz@ASone.c.u-tokyo.ac.jp} }

\inst{
Department of Basic Science, University of Tokyo, \\
3-8-1 Komaba, Meguro-City, Tokyo 153-8902, Japan \\[5pt]
$^{1}$Institute for International Advanced Interdisciplinary Research, \\
Tohoku University, Miyagi 980-8578, Japan
}
\abst{
We propose a new model suitable for a nonequilibrium molecular dynamics (MD) 
simulation of electrical conductors.
The model consists of classical electrons and atoms.
The atoms compose a lattice vibration system.
The electrons are scattered by electron-electron and electron-atom interactions.
Since the scattering cross section is physically more important 
than the functional form of a scattering potential, 
we propose to devise the electron-atom interaction potential in such a way 
that its scattering cross section agrees with that of quantum-mechanical one. 
To demonstrate advantages of the proposed model,
we perform a nonequilibrium MD simulation assuming 
a doped semiconductor at room or higher temperature. 
In the linear response regime, 
we confirm Ohm's law, the dispersion relations and the fluctuation-dissipation relation.
Furthermore, we obtain reasonable dependence of the electrical conductivity on temperature, despite the fact that our model is a classical model.
}

\kword{nonequilibrium steady state, nonequilibrium molecular dynamics simulation, 
electrical condution, doped semiconductors, lattice vibrations, scattering cross section, 
dispersion relation, fluctuation-dissipation relation, temperature dependence}

\begin{document}
\maketitle

\section{Introduction}
In statistical mechanics, 
construction of  nonequilibrium statistical mechanics has been attempted for long years.
Whereas in the linear nonequilibrium regime the linear response theory was established in the 1950s\cite{Kubo:1995}, 
in the nonlinear nonequilibrium regime the properties 
of nonequilibrium states far from equilibrium are still poorly understood.

When trying 
to investigate such nonequilibrium states with analytical approaches, 
one runs into difficulties of solving the equations of motion analytically. 
A promising approach is the nonequilibrium molecular dynamics (MD) simulation, 
which came into sight in 1950s.\cite{Alder:1957}
In this approach, 
assuming an appropriate microscopic model,
one numerically solves the equations of motion of all constituents, 
obtains the values of macroscopic variables of interest, 
and thereby understands properties of nonequilibrium states.

Transport phenomena are the most important in nonequilibrium 
statistical mechanics.
For heat conduction, many studies by the nonequilibrium MD simulations
were reported.
For example, the Fourier law was successfully obtained.\cite{Shimada:2000}
For charge or mass transport, in contrast, 
the nonequilibrium MD simulation
was successfully performed only recently
by Yuge, Ito and Shimizu in 2005.\cite{Yuge:2005} 
Their model, which we call the YIS model, 
consists of three types of classical particles,
which imitate  electrons, phonons, and impurities. 
These particles interact with each other via short-range interaction potentials.
Using the YIS model, 
it was subsequently shown for states far from equilibrium 
that response becomes strongly nonlinear \cite{Yuge:2005},
the long-time tail is significantly modified \cite{Yuge:2007,Yuge_alder:2009},
the fluctuation-dissipation relation (FDR) is significantly violated and universal excess noise
appears \cite{Yuge:2009}.
It was further shown that 
the sum rules and the asymptotic behaviors, 
which were recently derived in refs. \citen{ShimizuYuge:2010, Shimizu:2010, Yuge:2010}, 
of response functions of NESSs are indeed satisfied \cite{ShimizuYuge:2010}.

Although the YIS model have enjoyed such great successes in treating 
fundamental properties of NESSs,
it has two problems to treat more general properties.
First, 
we have found that 
the YIS model is not suitable for treating macroscopically inhomogeneous conductors \cite{Shimizu:1999},
because in NESSs of such conductors
the phonons are pushed by the electrons 
away from the conductors, which is physically unrealistic \cite{Lee:Mthesis}.
Second, the YIS model is inconvenient for
the analysis of the temperature dependence of the electrical conductivity,
because the number of phonons has to be changed 
as a function of temperature.
These difficulties arise because the phonons are treated as classical 
particles.

In this paper, we propose a new model which 
resolves these difficulties.
In this model, we represent a phonon system by a classical lattice vibration system. 
Each electron is scattered by lattice vibrations and by other electrons
(and possibly by impurities).
As an illustration of advantages of the present model, 
we will show  by MD simulations that 
not only nonequilibrium properties at each temperature
but also the temperature dependence of the conductivity 
can be analyzed in a natural way.

This paper is organized as follows.
We describe the essential elements to treat electrical conduction in 
Sec.~\ref{sec:EE}.
The new model, which includes all the essential elements,
is presented in Sec.~\ref{sec:model}. 
To illustrate advantages of the model, 
we perform an MD simulation in Sec.~\ref{sec:app} 
assuming a doped semiconductor at room or higher temperature. 
The results will be presented and discussed in Sec.~\ref{sec:results}.
We devote the last section to the summary.

%%%%%%%%%%%%%%%%%%%%%%%%%%%%%%%%%%%%%%%%%%%%%%%%%%%%%%%%%%%%%%%%%%%%%%%%%%%%%%%%%%%%%%%%%%%%%%%%%%%%%%%%%%%%%%%%%

\section{Essential Elements to Treat Electrical Conduction}
\label{sec:EE}

Before defining the model, 
we here present the essential elements to treat electrical conduction.

A typical experimental situation is as follows. 
An electrical conductor (e.g., a doped semiconductor) is put on a sample holder whose temperature is kept constant. 
The both ends of the conductor are connected to a battery through conducting wires. 
By inserting an ampere meter to the wire, one measures the conductance 
by the two- or four-terminal method. 
 
From a microscopic viewpoint, electrical conduction is explained as follows.
Energy is supplied to the conductor from the battery, 
and global motion of the electrons is induced, which results in a finite electrical current.
The supplied energy is transferred to the lattice through the electron-lattice interaction, 
and dissipates as the Joule heat into the sample holder through the lattice-bath coupling. 
A balanced state in which the supplied energy equals the Joule heat 
is a nonequilibrium steady state (NESS).

Therefore, a model for electrical conduction should have the following elements: 
(a) an interacting many-electron system, 
(b) an energy dissipating system, such as a lattice vibration system and a bath for the lattice, 
(c) objects violating the microscopic translational invariance 
to define the rest frame, such as impurities, fixed walls, and the lattice of atoms
(d) interactions among these constituents, and (e) an external force to drive electrons.

%%%%%%%%%%%%%%%%%%%%%%%%%%%%%%%%%%%%%%%%%%%%%%%%%%%%%%%%%%%%%%%%%%%%%%%%%%%%%%%%%%%%%%%%%%%%%%%%%%%%%%%%%%%%%%%%%%%%%%%%%%%%%%%%%%%%%%%%%%%%%%%%%%%%%%%%%%%%%%%

\section{Proposal of Model}
\label{sec:model}

We propose a model that includes all the elements in the previous section. 
From now on, we explain the following components one by one: 
a many-electron system, a lattice vibration system, 
a proper interaction between these two systems, impurities, 
an external electric field and the boundary conditions.

%%%%%%%%%%%%%%%%%%%%%%%%%%%%%%%%%%%%%%%%%%%%%%%%%%%%%%%%%%%%%%

\subsection{Many-Electron System}

We consider the regime where the electron temperature $k_{\rm{B}} T_{\rm{e}}$ is higher 
than the Fermi level $\epsilon_{\rm{F}}$ (the chemical potential at zero temperature): 
$k_{\rm{B}} T_{\rm{e}} \gtrsim \epsilon_{\rm{F}}$. 
In this regime, we can treat the electrons as classical particles.

The kinetic energy of the classical many-electron system is given by 
\begin{equation}
  K_{\rm{e}} = \sum_i \frac{m \Vec{v}_i^2}{2} \: ,
\end{equation}
where $m$ is the effective mass of an electron, 
and $\Vec{v}_i$ is the velocity of the electron labeled by $i$ 
($i = 1, 2, . . ., N_{\rm{e}}$).

We assume a conductor which is translation-invariant macroscopically. 
Then, the electron-electron interaction potential $U_{\rm{ee}}$ is reduced to a short-range potential, 
because long-range effects of the Coulomb potential are screened. 
It is expected that the detailed form of short-range potential $U_{\rm{ee}}$ has 
no significant effect on transport properties.
Hence, among many possible forms of short-range potentials, 
we here take the following simple one: 
\begin{equation}
 U_{\rm{ee}} = \sum_i \: \sum_{j > i} \: 
 Q_{\rm{ee}} \left( 1 - \frac{\vert \Vec{r}_{i} - \Vec{r}_{j} \vert}{l_{\rm{ee}}} \right)^4
 \Theta(l_{\rm{ee}} - \vert \Vec{r}_i - \Vec{r}_{j} \vert) \: ,
\end{equation}
where $Q_{\rm ee}$ is a positive constant, and
$\Vec{r}_i$ is the position of the $i$th electron, 
and $l_{\rm{ee}}$ is the interaction range, 
and $\Theta$ is the step function.
The factor $\Theta(l_{\rm{ee}} - \vert \Vec{r}_i - \Vec{r}_j \vert)$ 
ensures that pairs of electrons interact only when the electrons are within the distance $l_{\rm{ee}}$. 
The factor $(1 - {\vert \Vec{r}_i - \Vec{r}_j \vert}/{l_{\rm{ee}}})^4$ is 
introduced to weaken the singularity of $U_{{\rm ee}}$ at $\vert \Vec{r}_i - \Vec{r}_j \vert = l_{{\rm ee}}$. 
When the constant $Q_{\rm{ee}}$ is much larger than $k_B T_{\rm{e}}$, 
$U_{\rm{ee}}$ can be regarded as a hard core potential.

%%%%%%%%%%%%%%%%%%%%%%%%%%%%%%%%%%%%%%%%%%%%%%%%%%%%%%%%%%%%%%%%%%%%%%%%%%%%%%%%%%%%%%%%%%%%%%%%%%%%%%%%%%%%%%%%%%%%%%%

\subsection{Lattice Vibration System}

We consider the case where the temperature is higher than 
the Debye frequency: $k_{\rm{B}} T \gtrsim \hbar \omega_{\rm{D}}$. 
In this case we can treat the quantum lattice vibration system as a classical lattice vibration system. 
It consists of many classical atoms, whose kinetic energy is given by 
\begin{equation}
  K_{\rm{a}} = \sum_\alpha \frac{M \Vec{V}_\alpha^2}{2} \: ,
\end{equation}
where $M$ is the mass of an atom, and $\Vec{V}_\alpha$ is the velocity of the atom labeled by $\alpha$ 
($\alpha = 1, 2, . . ., N_{\rm{a}}$).

Suppose that the atoms compose a square lattice, 
whose lattice constant at mechanical equilibrium is denoted by $l$. 
We assume that each atom is connected to its nearest-neighbor atoms by nonlinear springs. 
The atom-atom interaction potential $U_{\rm aa}$ is taken as 
\begin{equation}
 U_{\rm{aa}} = \sum_{\langle \alpha  \, \beta \rangle} \left( { \frac{K}{2} (\vert \Vec{R}_{\alpha} - \Vec{R}_{\beta} \vert - l_0) ^2 
                                                 + \frac{G}{4} (\vert \Vec{R}_{\alpha} - \Vec{R}_{\beta} \vert - l_0) ^4} \right) \: ,
\end{equation}
where the summation is taken over all pairs of nearest neighbors. 
$\Vec{R}_\alpha$ is the position of the $\alpha$th atom, $l_0$ is the natural length of each spring, 
and $K$ and $G$ are positive constants.
Since the nonlinear terms are included in $U_{\rm{aa}}$,  
the lattice system is expected to be chaotic 
if $N_{\rm a}$ is large enough.
It is therefore expected to have good statistical properties.
Furthermore we impose the condition 
\begin{equation}
  l > l_0 \: ,
\end{equation}
to prevent the zero-frequency angular modes. 

We also assume the following self potential: 
\begin{equation}
 U_{\rm{a}} =  \sum_{\alpha} \left( { \frac{K'}{2} \vert \Vec{R}_{\alpha} - \Vec{R}_{\alpha_0} \vert ^2 
                                    + \frac{G'}{4} \vert \Vec{R}_{\alpha} - \Vec{R}_{\alpha_0} \vert ^4 } \right) \: ,
\end{equation}
where $K'$ and $G'$ are positive constants, 
and $\Vec{R}_{\alpha0}$ represents the mechanical equilibrium position of the $\alpha$th atom.
This potential stabilizes the positions of the atoms around their mechanical equilibrium positions more rigidly.

We impose the condition that the lattice doesn't melt; 
\begin{equation}
 l \gg \delta \: ,
 \label{melt}
\end{equation} 
where $\delta$ is the standard deviation of the position of an atom: 
$\delta \equiv \sqrt{\langle |\Vec{R}_\alpha - \Vec{R}_{\alpha 0}|^2 \rangle}$
($\langle A \rangle$ denotes the averaged value of $A$), 
and $\delta \simeq \sqrt{k_{\rm{B}} T/K}$ when $G = K' = G' = 0$. 

We note that the mass of an electron is much smaller than that of an atom; 
\begin{equation}
  m \ll M \: .
\end{equation}
Hence, even if momentum of an electron ($\sim \sqrt{k_{\rm{B}} T_{\rm{e}} m}$) 
is fully transferred to an atom, its effect on the vibrating motion of the atom is quite small.

%%%%%%%%%%%%%%%%%%%%%%%%%%%%%%%%%%%%%%%%%%%%%%%%%%%%%%%%%%%%%%%%%%%%%%%%%%%%%%%%%%%%%%%%%%%%%%%%%%%%%%%%%%%%%%%%%%%%%%%%

\subsection{Interaction between Many-Electron System and Lattice Vibration System}

In this subsection, we propose a \textit{classical-mechanical} 
form of the electron-atom interaction potential $U_{\rm{ea}}$, 
which reproduces a \textit{quantum-mechanical} scattering cross section. 

\subsubsection{Electron-phonon scattering in a quantum-mechanical system}

Before discussing a classical form, we briefly review 
the theory of electron-phonon scattering in solids.\cite{Bardeen:1950}\cite{Ashcroft:1976}

Consider the single-electron potential produced by the lattice. 
When the lattice is perfectly periodic, 
the potential (denoted by $\hat{U}_0(\Vec{r})$) is also periodic: 
$\hat{U}_0(\Vec{r}) = \hat{U}_0(\Vec{r} + \Vec{n})$, 
where $\Vec{n}$ is the lattice vector. 
In this case we can incorporate the effect of the lattice potential on an electron 
by replacing the electron momentum with the crystal momentum, 
and by modifying the electron's dispersion relation. 
When the electron energy is much smaller than the band width, 
we can employ the effective-mass approximation. 

When the positions of the atoms are displaced from the periodic structure, 
the lattice potential (denoted by $\hat{U}_d(\Vec{r})$) is non-periodic, 
and the difference, $\hat{V} = \hat{U}_d(\Vec{r}) - \hat{U}_0(\Vec{r})$, causes the electron-phonon scattering.

%and the Schr\"{o}dinger equation 
% $\left( - \hbar^2 \nabla^2/2m_{e}  + U_0(\Vec{r}) \right) \psi(\Vec{r}, \Vec{k})
% = E_0(\Vec{k}) \psi(\Vec{r}, \Vec{k})$
%holds where $m_{e}$ is the mass of an electron, 
%the wave function of an electron could be represented as
%$\psi(\Vec{r}, \Vec{k}) = \exp \left( i \Vec{k} \cdot \Vec{r} \right) u(\Vec{r}, \Vec{k})$
%where $u(\Vec{r}, \Vec{k}) = u(\Vec{r} + \Vec{T}, \Vec{k})$ holds.
%Therefore, the momentum could be reduced to the crystal momentum $\hbar \Vec{k}$ 
%and electron is ineffectively scattered by a perfect lattice $U_0(\Vec{r})$. 
%If $|\Vec{k}|$ is smaller than the absolute value of the reciprocal lattice $|\Vec{K}|$ 
%(for example in a doped semiconductor, this condition is satisfied.), 
%the energy $E_0(\Vec{k})$ could be reduced to $E_0(\Vec{k}) = E_0(0) + \hbar^2 \Vec{k}/2 m$ 
%where $m$ is the effective mass of an electron $m\equiv \hbar^2 (\lim_{\vert \V%ec{k} \vert \rightarrow 0} (\partial^2 E_0(\Vec{k})/\partial \Vec{k})^2)^{-1}$.%In short, 

The scattering cross section $s_{\textrm{quantum}}$ 
in the Born approximation is proportional to $|\langle f|\hat{V}|i \rangle|^2$ 
where $|i\rangle$ and $|f\rangle$ are the states before and after the scattering, respectively. 
Its magnitude depends on the phonon mode. 

For the scattering by the acoustic modes in the long wave length limit, 
for example, $\hat{V} \simeq E_l \nabla \cdot \Vec{d}(\Vec{r})$, 
where $E_l$ is a constant, and $\Vec{d}$ is the displacement of an atom 
from its position on the perfect lattice. 
In the Fourier space, $\hat{V}_{\Vec{q}} \propto \Vec{q} \cdot \Vec{d}(\Vec{q})$. 
Therefore $s_{\textrm{quantum}} \propto |\langle f | V | i \rangle|^2 
\propto |\nabla \cdot \Vec{d}(\Vec{r})|^2 \propto |\Vec{d}_\parallel(\Vec{q})|^2$, 
where $\Vec{d}_\parallel(\Vec{q})$ denotes 
the component of $\Vec{d}(\Vec{q})$ parallel to $\Vec{q}$.

\subsubsection{Electron-atom scattering potential in the classical-mechanical model}
\label{Electron-atom scattering potential in the classical-mechanical model}

Since we are considering the case where $k_{\rm{B}} T_{\rm{e}} \gtrsim \epsilon_{\rm{F}}$, 
quantum interference effects are weak, and the electrons may be treated as classical particles. 
Let us consider the scattering potential $U_{\rm{ea}}$ for the classical electrons. 

Generally, to model a quantum-mechanical system using a classical-mechanical system, 
the scattering cross section is much more important than the functional form of a scattering potential 
\cite{Shimizu:unpublished}.

In the present case, if one employed that the classical potential $U_{\rm{ea}}$ 
that has exactly the same form 
as the quantum potential $\hat{U}_{\rm{ea}}$, 
then the classical electrons would be scattered by $U_{\rm{ea}}$ 
even when the atoms are not displaced from the lattice points. 
This effect sharply contradicts the nature of the quantum system, 
in which electrons are not scattered (except for Umklapp processes) by a perfect lattice of the atoms. 

We therefore propose that $U_{\rm{ea}}$ should be devised in such a way that 
the classical scattering cross section $s_{\textrm{classical}}$ 
imitates 
the quantum scattering cross section $s_{\textrm{quantum}}$ well.
Specifically 
when we consider the case where 
$s_{\textrm{quantum}} 
\propto |\Vec{d}|^2$, 
$U_{\rm{ea}}$ should lead to 
$s_{\textrm{classical}} \propto \delta^2$ 
for each atom.
If $U_{\rm ea}$ satisfies this condition, 
it is expected that the detailed form of $U_{\rm ea}$ is irrelevant to transport properties. 

As an example of such $U_{\rm{ea}}$ in a two dimensional system, 
we here take \cite{appendix} 
\begin{eqnarray}
U_{\rm{ea}}
&=& 
\sum_\alpha \sum_\beta 
Q_{\rm{ea}} 
\left| 
\frac{\Vec{R}_\beta - \Vec{R}_{\beta 0}}{l} 
\right|^8
\left( 
1 - \frac{ l \: |\Vec{r}_\alpha - \Vec{R}_\beta| }
{ \gamma_{\rm{ea}} | \Vec{R}_\beta - \Vec{R}_{\beta 0} |^2 } 
\right)^4 
\nonumber\\
& & \quad \times \
\Theta \left( 
\gamma_{\rm{ea}} \left|
{ \frac{\Vec{R}_\beta - \Vec{R}_{\beta 0}}{l}} 
\right|^2 
- \left| {\frac{\Vec{r}_\alpha - \Vec{R}_\beta}{l}} \right|
\right),
\label{Uea_2d}
\end{eqnarray}
where $Q_{\rm{ea}}$ and $\gamma_{\rm{ea}}$ are
positive constants which characterize the magnitude and 
the interaction range
of $U_{\rm{ea}}$, respectively.
Due to the step function $\Theta$, 
the interaction range (for an electron) of this potential is 
$\sim \gamma_{\rm{ea}} |\Vec{R}_\beta - \Vec{R}_{\beta 0}|^2 / l$ 
($l$ is the lattice constant),
which is proportional to the square of the displacement 
$|\Vec{R}_\beta - \Vec{R}_{\beta 0}|$ of the atom.
Since $s_{\textrm{classical}}$
is proportional to the interaction range in a two dimensional system, 
we have 
$s_{\textrm{classical}} \propto \delta^2$, 
as required.
The factor 
$\left( 
1 - 
l |\Vec{r}_\alpha - \Vec{R}_\beta| /
\gamma_{\rm{ea}} | \Vec{R}_\beta - \Vec{R}_{\beta 0} |^2  
\right)^4 
$
is introduced to weaken the singularity at 
$
|\Vec{r}_\alpha - \Vec{R}_\beta| 
=
\gamma_{\rm{ea}} | \Vec{R}_\beta - \Vec{R}_{\beta 0} |^2/l
$. 
Although this factor is singular with respect to $\Vec{R}_\beta$
at $\Vec{R}_\beta = \Vec{R}_{\beta 0}$, 
this singularity is removed by the extra multiplicative factor 
$
\left| 
\Vec{R}_\beta - \Vec{R}_{\beta 0}/l 
\right|^8
$. 

For a three dimensional system, 
$s_{\textrm{classical}}$
is proportional to the {\em square} of the interaction range. 
We can therefore take $U_{\rm{ea}}$, for example, as
\begin{eqnarray}
U_{\rm{ea}}
&=& 
\sum_\alpha \sum_\beta 
Q_{\rm{ea}} 
\left| 
\frac{\Vec{R}_\beta - \Vec{R}_{\beta 0}}{l} 
\right|^4
\left( 
1 - \frac{ |\Vec{r}_\alpha - \Vec{R}_\beta| }
{ \gamma_{\rm{ea}} | \Vec{R}_\beta - \Vec{R}_{\beta 0} | } 
\right)^4 
\nonumber\\
& & \quad \times \
\Theta \left( 
\gamma_{\rm{ea}} \left|
{ \frac{\Vec{R}_\beta - \Vec{R}_{\beta 0}}{l}} 
\right| 
- \left| {\frac{\Vec{r}_\alpha - \Vec{R}_\beta}{l}} \right|
\right).
\label{Uea_3d}
\end{eqnarray}
This gives $s_{\textrm{classical}} \propto \delta^2$, as required, in a three dimensional system.

%%%%%%%%%%%%%%%%%%%%%%%%%%%%%%%%%%%%%%%%%%%%%%%%%%%%%%%%%%%%%%%%%%%%%%%%%%%%%%%%

\subsection{Impurities}

To violate the microscopic translational invariance and thereby 
define the rest frame of equilibrium states, we add a random potential. 
We represent the random potential as the sum of short-range potentials produced by impurities.
The impurities are modeled by particles which are fixed at random positions and interact with 
the electrons and the atoms through the short-range potentials. 
Among many possible forms of the short-range potentials, 
we here take the following forms for 
the electron-impurity potential $U_{\rm{ei}}$ 
and the atom-impurity potential $U_{\rm{ai}}$: 
\begin{equation}
 U_{\rm{ei}} = \sum_{i} \: \sum_{k} \: 
 Q_{\rm{ei}} \left( 1 - \frac{\vert \Vec{r}_{i} - \Vec{\xi}_{k} \vert}{l_{\rm{ei}}} \right)^4
 \Theta(l_{\rm{ei}} - \vert \Vec{r}_i - \Vec{\xi}_k \vert) \: ,
\end{equation}
\begin{equation}
 U_{\rm{ai}} = \sum_{\alpha} \: \sum_k \: 
 Q_{\rm{ai}} \left( 1 - \frac{\vert \Vec{R}_{\alpha} - \Vec{\xi}_k \vert}{l_{\rm{ai}}} \right)^4
 \Theta(l_{\rm{ai}} - \vert \Vec{R}_\alpha - \Vec{\xi}_k \vert) \: ,
\end{equation}
where $\Vec{\xi}_k$ is the position of the $k$th impurity ($k = 1, 2, ..., N_{\rm i}$). 
$Q_{\rm{ei}}$ and $Q_{\rm{ai}}$ are the positive constants. 
$l_{\rm{ei}}$ and $l_{\rm{ai}}$ are the interaction ranges. 

%%%%%%%%%%%%%%%%%%%%%%%%%%%%%%%%%%%%%%%%%%%%%%%%%%%%%%%%%%%%%%%%%%%%%%%%%%%%%%%%%%%%%%%%%%%%%%%%%%%%%%%%%%%%%%%%%%%%%%%%%

\subsection{Boundary Conditions}

To sum up, we have obtained the following Hamiltonian:
\begin{equation}
 H = K_{\rm{e}} + K_{\rm{a}} + U_{\rm{ee}} + U_{\rm{ea}} + U_{\rm{aa}} + U_{\rm{a}} + U_{\rm{ei}} + U_{\rm{ai}} \: .
\end{equation}
In addition, an electric field in the $x$-direction $E$ is applied to all electrons, 
and each electron experiences a force $F = e E$.

\begin{figure}[htbp]
  \centering
  \includegraphics[clip,width=.85\linewidth]{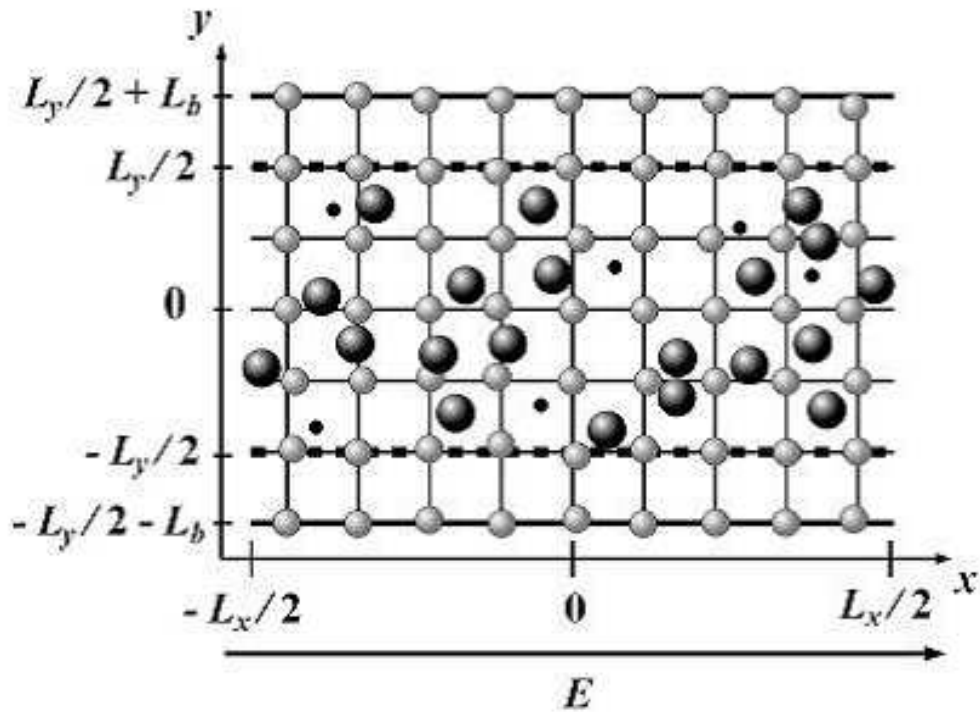}
  \caption{A schematic picture of the two-dimensional model of a macroscopically homogeneous conductor.
  	   The large black circles and the small black circles represent electrons and impurities, 
           respectively. The gray circles represent atoms, which compose a square lattice. 
           By an electric field $E$, the electrons 
           move from the left to the right in average, interacting with the atoms and the impurities.
  \label{uniform_system_diagram}}
\end{figure}

Hereafter, we consider two-dimensional systems, 
employing eq.~(\ref{Uea_2d}) in 
\S \textit{\ref{Electron-atom scattering potential in the classical-mechanical model}}
as the electron-atom scattering potential. 
A schematic picture of the model is shown in Fig.~\ref{uniform_system_diagram}. 

The boundary conditions are imposed as follows.
The boundaries in the $x$-direction are the periodic boundaries for all particles.
The boundaries in the $y$-direction for electrons are potential walls, 
which simulate the walls at the boundaries of the conductor.
Those for atoms are thermal walls, which are located 
away from the potential walls for electrons by a distance $L_{b}$ (see Fig.~\ref{uniform_system_diagram}). 
The thermal walls simulate the thermal contact with a sample holder whose temperature is kept at $T$. 
Through these thermal walls, heat is transferred outside the conductor, 
and a NESS can be realized in the conductor. 

Note that the present model is a pure mechanical model, except for the thermal walls for atoms. 

%%%%%%%%%%%%%%%%%%%%%%%%%%%%%%%%%%%%%%%%%%%%%%%%%%%%%%%%%%%%%%%%%%%%%%%%%%%%%%%%%%%%%%%%%%%%%%%%%%%%%%%%%%%%%%%%%%%%%

\section{Application to Doped Semiconductors}\label{sec:app}

To illustrate advantages of the proposed model, 
we apply the model to doped semiconductors, and perform an MD simulation.

\subsection{Doped Semiconductors}

The proposed model can be applied 
to doped semiconductors at room or higher temperature 
(but lower than the melting temperature). 
We explicitly show this for the case where the electrons are confined 
in a two-dimensional plane (e.g., in a quantum well or inversion layer) 
by showing that the conditions of Sec. \ref{sec:model} are satisfied. 

Since the electron temperature $T_{\rm {e}}$ in NESSs is higher than the bath temperature $T$, 
$k_{\rm{B}} T_{\rm{e}} \geq k_{\rm{B}} T \geq k_{\rm{B}} T_{\textrm{room}} \simeq 26\textrm{meV}$. 
On the other hand, $\epsilon_{\rm{F}} \simeq \pi \hbar^2 n_{\rm{e}}/m$ 
where $n_{\rm{e}}$ is the electron density. 
For $n_{\rm e} \simeq 5.0 \times 10^{15}\textrm{m}^{-2}$, 
for example, 
$\epsilon_{\rm{F}} \simeq 6.5\textrm{meV}$ in doped silicon (Si) semiconductors, 
whereas 
$\epsilon_{\rm{F}} \simeq 18\textrm{meV}$ in doped gallium arsenide (GaAs) semiconductors. 
In both cases, the condition $k_{\rm{B}} T_{\rm{e}} \gtrsim \epsilon_{\rm{F}}$ is satisfied.

The condition $k_{\rm{B}} T \gtrsim \hbar \omega_{\rm{D}}$ is approximately satisfied, 
because in Si and GaAs crystals, the Debye temperature is 645K and 633K, respectively; 
hence $k_{\rm{B}} T \sim \hbar \omega_{\rm{D}}$.

The condition $l \gg \delta \sim \sqrt{k_B T/K}$ is also satisfied, 
because we are considering the case where the temperature is less than the melting temperature. 
(In {\rm Si} and in {\rm GaAS} crystals, the melting temperature is 1687{\rm K} and 1511{\rm K}, respectively.)

Generally, at room or higher temperature, 
electron-atom scattering occurs more frequently than electron-impurity scattering. 
Hence, we will take the model parameters in such a way that 
\begin{equation}
  \tau_{\rm{ea}} \ll \tau_{\rm{ei}}
\end{equation}
is satisfied, where ${\tau_{\rm{ea}}}$ is the mean free time between electron-atom collisions, 
and ${\tau_{\rm{ei}}}$ is that between electron-impurity collisions. 

Therefore, our model is applicable to typical doped semiconductors. 

\subsection{MD Simulation}

We analyze electrical conduction in doped semiconductors by an MD simulation 
of the proposed model.
We set the lattice constant $l$, the effective mass of an electron $m$, 
the electric charge $e$, and a certain reference energy to unity. 

As illustrated in Fig.~\ref{uniform_system_diagram}, 
electrons and impurities are confined in $-L_x/2 \leq x \leq L_x/2$ and $-L_y/2 \leq y \leq L_y/2$, 
whereas atoms are confined in $-L_x/2 \leq X \leq L_x/2$ and $-(L_y/2 + L_b) \leq Y \leq L_y/2 + L_b$. 

At an initial time ($t = 0$), we put atoms at their mechanical equilibrium positions, 
and arrange electrons and impurities at random positions 
so as not to contact with each other. 
The initial velocities of electrons and atoms are given by Maxwell distribution of the bath temperature $T$. 
In other words, the initial state is closed to an equilibrium state.

We set the numbers of electrons ($N_{\rm e}$), atoms ($N_{\rm a}$), and impurities ($N_{\rm i}$) 
by considering the density of a real doped semiconductor.
In our simulation, we take the density of electrons high ($N_{\rm{e}} / N_{\rm{a}} \simeq 1 / 3$) 
to shorten the computational time, 
although in a usual doped semiconductor the density of electrons are rather low ($N_{\rm{e}} / N_{\rm{a}} \ll 1$). 

An external electric field $E$ is applied for $t > 0$. 
Electrons and atoms move according to Newton's equations of motion.
We use Gear's fifth-order predictor-corrector method to solve the equations of motion. 
We also use the Linked Cell method\cite{Quentrec:1975} 
in calculating the force due to $U_{\rm{ee}}$, $U_{\rm{ea}}$, $U_{\rm{ei}}$ and $U_{\rm{ai}}$. 

%If we calculated the force of all pairs, the calculation time would be $O$($N^2$), 
%and which would be unsuitable for a large-scale simulation.
%Therefore, we use the Linked Cell method
%for 
%This method takes $O$($N$) calculation time.
%Although $M / m \gtrsim 10^4$ in the real systems, 
%we take $M = 10$ ($m = 1$) to shorten the relaxation time.

We set $k_{\rm{B}}T = 1$, except when we investigate the temperature dependence. 
We set the parameters as follows: 
$Q_{\rm{ee}} = 10000$, $l_{\rm{ee}} = 1/2$, 
%(i.e., the electrons interact with each other when they overlap), 
$M = 10$, $Q_{\rm{ea}} = 10^{10}$, $\gamma_{\rm{ea}} = 1$, 
%(another choice is to take the magnituide of the potential $Q_{\rm ee}$, $Q_{\rm ea}$ smaller, 
%and to take $l_{\rm ee}$, $\gamma_{\rm ea}$ larger), 
$Q_{\rm{ei}} = 10000$, $l_{\rm{ei}} = 1/12$, 
%(i.e., the electrons are scattered when their centers are within the radius of the atom or the impurity), 
$Q_{\rm{ai}} = 10000$, $l_{\rm{ai}} = 1/4$, 
%(i.e., the atoms and the impurities interact with each other when they overlap),  
$K = 100$, $G = 800$, $K' = 100$, $G' = 800$ and $l_0 = 3/4$. 
Here, we have taken 
$Q_{\rm ea} \left( \delta/l \right)^8 > k_{\rm B} T_{\rm e} \sim k_{\rm B} T$ 
to estimate $\tau_{\rm ea}$ easily, and 
$l_{\rm ei}$ small not to impede the electrons' flow too much.
For these values of the parameters, 
effects of the nonlinear terms are small, 
because $K \delta^2 / 2 \gg G \delta^4 / 4$ and $K' \delta^2 / 2 \gg G' \delta^4 / 4$.
These parameters satisfy the conditions in the previous sections: 
$l > l_0$, $l \gg \delta$, $M \gg m$, and $\tau_{\rm ea} \ll \tau_{\rm ei}$ 
(for the particle densities assumed in the following MD simulations).

In the $x$-direction, we take the periodic boundary condition.
In the $y$-direction for electrons, the boundaries are the potential walls.
The wall potential is taken as $U = Q_{\rm{ee}} (w/l_{\rm{ee}})^4$, 
where $w$ is the penetration depth of an electron into the wall.
For atoms, the end atoms around $|y| = L_y / 2 + L_b$ are connected to hard walls 
which are located at a distance $l$ away from the mechanical equilibrium positions of the end atoms 
(i.e., at $|y| = L_y / 2 + L_b + l$). 
When an end atom comes in contact with a thermal wall 
(located at $|y| = L_y / 2 + L_b$), \cite{Hatano:1998}
the atom loses its memory and is reflected back with a random velocity\cite{Tehver:1998}
whose probability distribution function is given by
\begin{equation}
 f(\Vec{V}) = \frac{1}{\sqrt{2 \pi}} \left( \frac{M}{k_{\rm{B}} T} \right)^{3/2} |V_y|
 \exp \left( - \frac{M\Vec{V}^2}{2k_{\rm{B}} T} \right) \: .
\end{equation}

In the main simulations, we fix the time-step width to be $10^{-4}$. 
We have confirmed that macroscopic behaviors of the system 
are almost independent of the time-step width 
if the width is taken smaller than this value.

We calculate the electrical current as follows. 
The $x$-component of the electrical current density at time $t$ 
is given by $j(x, y, t) \equiv \sum_{i} e \: v_i^x(t) \: \delta(x-x_i) \delta(y-y_i)$, 
where $v_i^x(t)$ denotes the $x$-component of the velocity of the $i$th electron. 
The current in the $x$ direction at $x$ is given by 
$I(x, t) \equiv \int dy \: j(x, y, t) = \sum_{i} \: e \: v_i^x(t) \delta(x - x_i)$. 
When measuring the current in an experiment, 
one usually measures the current averaged over some finite length $L$ in the $x$-direction: 
$I(t) = \int dx \: I(x, t) / L
        = e \sum_{i} v_i^x(t) / L \:$.
In this simulation, we take $L = L_x$, i.e., 
\begin{equation}
  I(t) = e \: \sum_{i \in \textrm{conductor}} \frac{v_i^x(t)}{L_x} \: \: .
\end{equation}

%%%%%%%%%%%%%%%%%%%%%%%%%%%%%%%%%%%%%%%%%%%%%%%%%%%%%%%%%%%%%%%%%%%%%%%%%%%%%%%%%%%%%%%%%%%%%%%%%%%%

\section{Results of MD simulation}
\label{sec:results}

\subsection{Snapshot}

\begin{figure}[htbp]
\centering
   \includegraphics[clip,width=.9\linewidth]{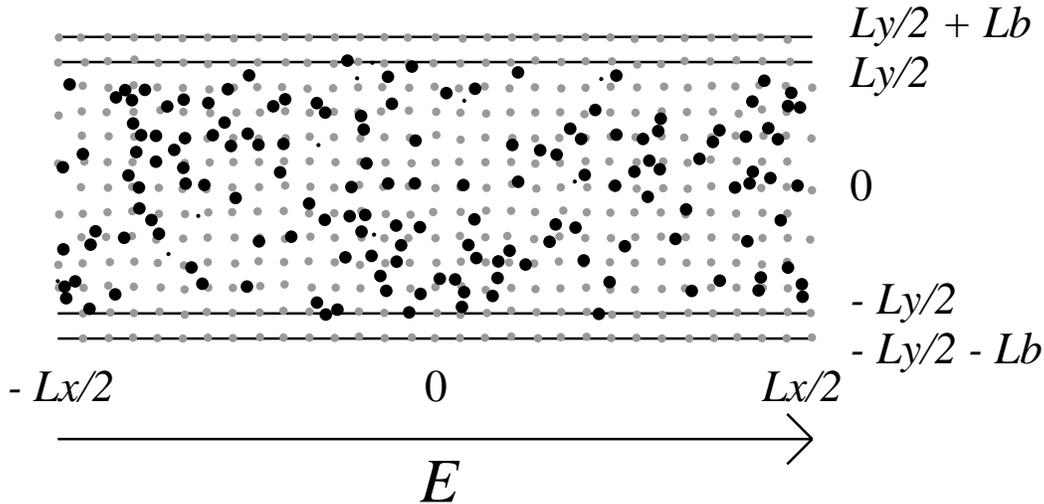}
   \caption{
   A snapshot of a NESS in a conductor at $E = 0.005$. 
   The large black circles, the gray circles, and the small black circles 
   represent the typical interaction ranges of 
   electrons, atoms, and impurities, respectively. 
   \label{uniform_system_snapshot}}
\end{figure}

Figure \ref{uniform_system_snapshot} is a snapshot of a NESS at $E = 0.005$ 
for a small-size system with $L_x = 30, L_y = 10, L_b = 1, N_{\rm{e}} = 150, N_{\rm a} = 390, N_{\rm{i}} = 10$. 
We observe that atoms (gray circles) compose a well-defined lattice, 
whereas electrons (large black circles) distribute rather randomly. 
Similar snapshots (not shown) are obtained for larger-size systems used in the following simulations.

\subsection{Realization of a NESS}

We first investigate whether the system reaches a NESS in the presence of an external electric field. 
A NESS is said to be realized if every macroscopic variable fluctuates around a constant value
and if the fluctuation is relatively small for a large-size system\cite{ShimizuYuge:2010, Shimizu:2010}.
We here examine two macroscopic variables: 
the electrical current in the $x$-direction, $I(t)$, 
and the total kinetic energy, $K(t)$.
To smear out high-frequency components, 
which will make the figures too busy,
we plot $I(t)$ and $K(t)$ after averaging over time intervals $\Delta t = 6$.

\begin{figure}[htbp]
  \centering
  \includegraphics[clip,width=.85\linewidth]{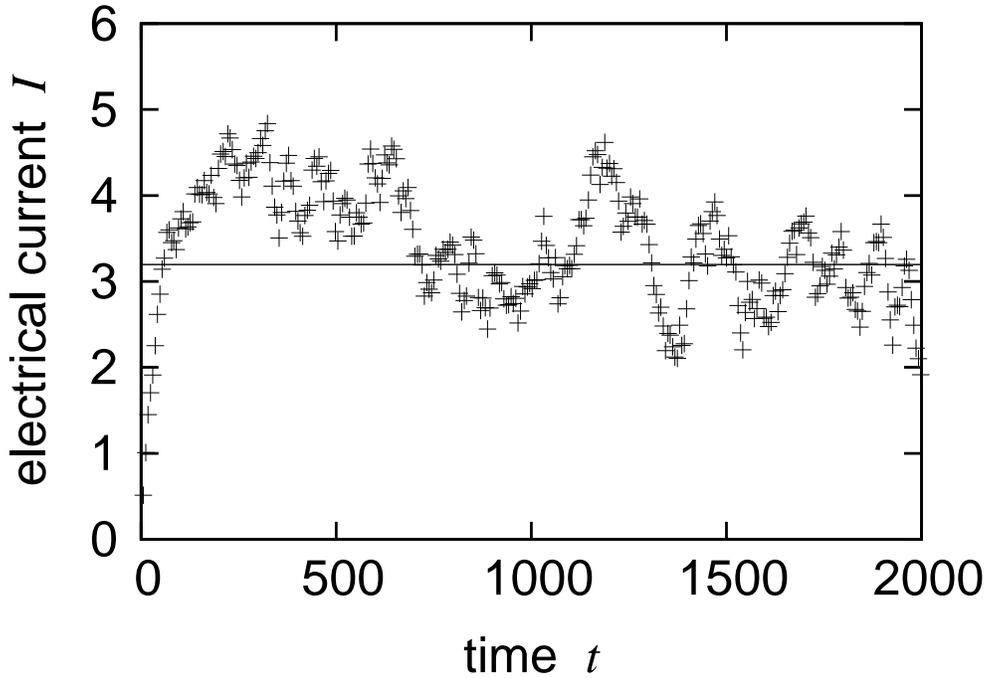}
  \caption{Time evolution of the electrical current $I(t)$ at $E = 0.005$.
           The solid line shows the time-averaged value of $I(t)$ during $1000 < t \leq 2000$. \label{t_I}}
\end{figure}

Figure \ref{t_I} depicts the time evolution of $I(t)$ for $E = 0.005$ 
for a large-size system with $L_x = 100, L_y = 36, L_b = 2, N_{\rm{e}} = 1500, N_{\rm a} = 4100, N_{\rm{i}} = 100$. 
For $t \gtrsim 800$, we see that 
$I(t)$ fluctuates around the long-time-averaged value 
(represented by the solid line in the figure). 
Although the fluctuation might look rather large,
its magnitude is reasonable for this size of system.
In fact, the fluctuation is of the same order of magnitude as 
that in an equilibrium state ($E=0$), 
and the latter is consistent with the fluctuation-dissipation 
relation as will be shown in \S \textit{\ref{sec:FDR}}.

\begin{figure}[htbp]
  \centering
  \includegraphics[clip,width=.85\linewidth]{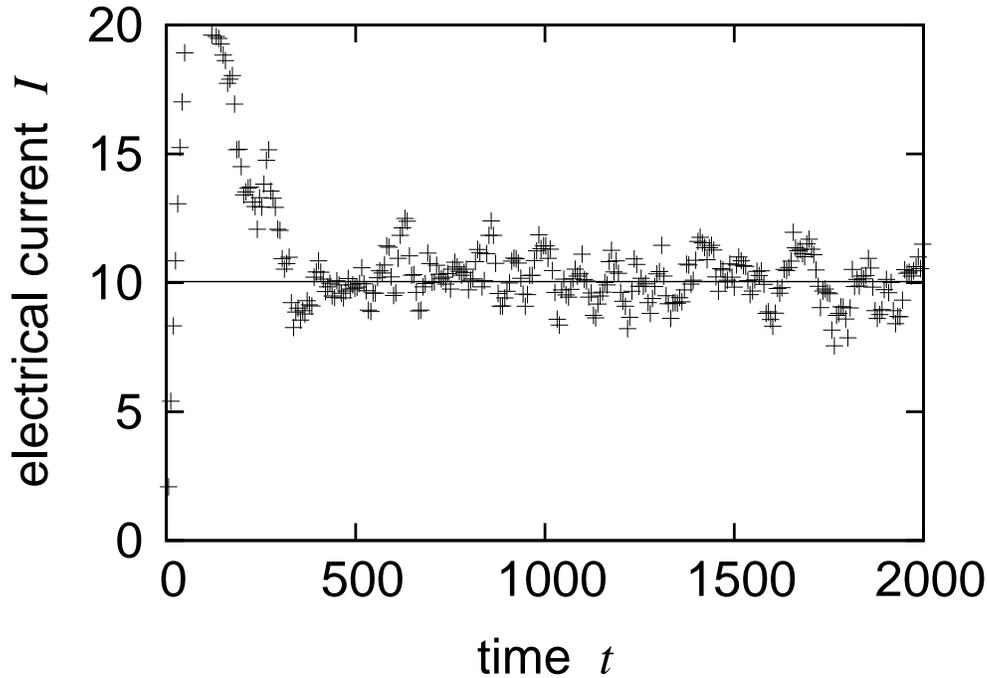}
  \caption{Time evolution of the electrical current $I(t)$ at $E = 0.04$.
           The solid line shows the time-averaged value of $I(t)$ during $1000 < t \leq 2000$. \label{t_I_v400}}
\end{figure}

Figure \ref{t_I_v400} depicts the time evolution of $I(t)$ for larger $E$ ($E = 0.04$). 
Comparing Figs. \ref{t_I} and \ref{t_I_v400}, 
we see that the \textit{relative} fluctuation for large $E$ is smaller 
than that for small $E$. 

\begin{figure}[htbp]
  \centering
  \includegraphics[clip,width=.85\linewidth]{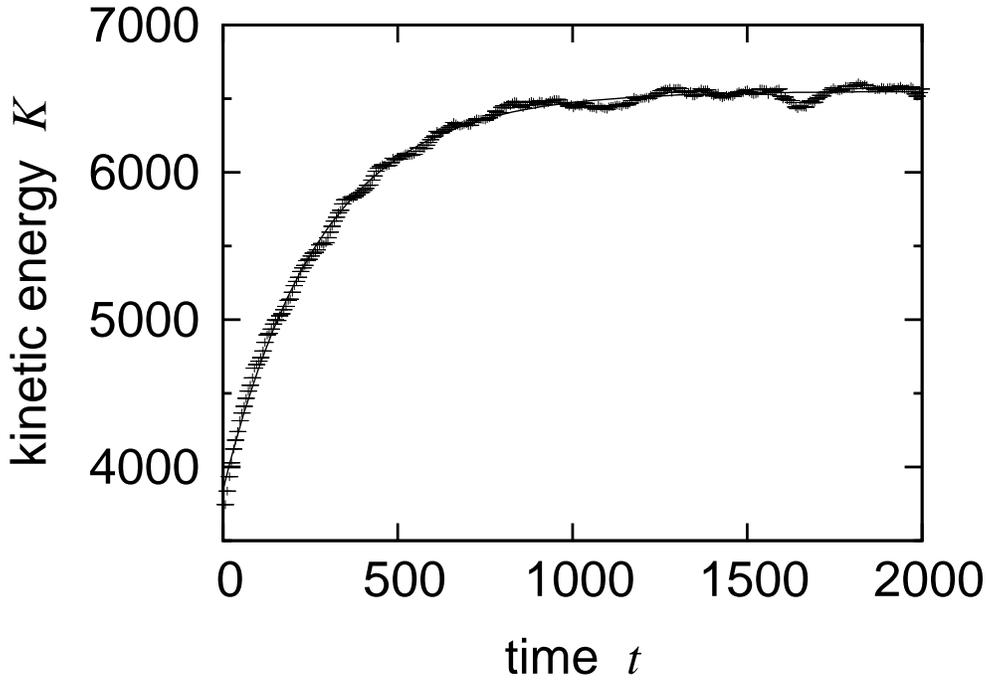}
  \caption{Time evolution of the kinetic energy $K(t)$ 
           as a function of time at $E = 0.005$. 
           The solid line is the fitting curve 
           whose functional form is given by eq. (\ref{K(t)}). \label{t_K}}
\end{figure}

Figure \ref{t_K} depicts the time evolution of 
$K(t) \equiv K_{\rm{e}}(t) + K_{\rm{a}}(t)$ for $E = 0.005$,
where $K_{\rm{e}}$ and $K_{\rm{a}}$ are the kinetic energies of 
electrons and atoms, respectively. 
For $t \gtrsim 1000$, 
$K(t)$ fluctuates around the long-time-averaged value, and 
the fluctuation is small enough.
Furthermore, 
$K(t)$ is fitted well to the function, 
\begin{equation}
K(t) = K(0) + a(1 - e^{-t/\tau_{\rm relax}}), \label{K(t)}
\end{equation}
which is represented by the solid line in the figure, 
where $K(0) = 3800$, $a = 2700$, $\tau_{\textrm{relax}} = 280$. 
This indicates that 
influence of the initial state becomes very small 
for $t \gtrsim 1000$ (because $\exp(-1000/280) \simeq 0.028$).

From these observations, we conclude that the system reaches a NESS
in the presence of an external electric field after a sufficiently long time. 
Hence, in the analysis below, 
we will use the time-averaged values calculated for $t > 1000$
for the average values of macroscopic variables in a NESS.

\subsection{Dependence on Electric Field}

Next, we investigate the dependence of the electrical current $I$ on the electric field $E$. 
We calculate long-time averaged values of $I$ independently for five systems with different impurity configurations, 
and calculate the average and standard deviation of the long-time averaged values. 
This is because we want to investigate the conductivity 
$\sigma = I / L_y E$ in a sufficiently large-size system. 
In an MD simulation, we can calculate the conductivity only in an insufficiently large-size system. 
We therefore take five impurity configurations, and estimate the conductivity in a sufficiently large-size system. 

\begin{figure}[htbp]
   \centering
   \includegraphics[clip,width=.85\linewidth]{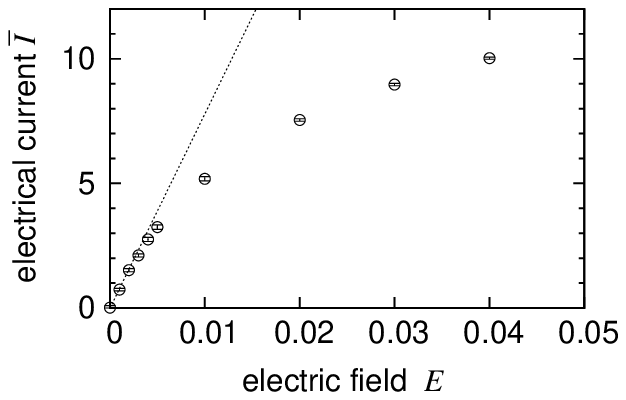}
   \caption{Average values of the electrical current $I$ plotted against the electric field $E$. 
            The error bars indicate the standard deviation over five impurity configurations. 
            The dotted curve represents the linear response. \label{E_I}}
 \end{figure}

The result is shown in Fig.~\ref{E_I}, 
where the long time averaged value of $I(t)$, denoted by $\overline{I}$, is plotted against $E$. 
We see that conductivity is nearly constant for $E \lesssim 0.003$. 
Therefore, Ohm's law $\overline{I} \propto E$ holds for small $E$ ($E \lesssim 0.003$), 
whereas the nonlinear response is observed for larger $E$. 
The standard deviations are very small, which implies that the dependence 
on the impurity configuration is relatively small. 

\begin{figure}[htbp]
   \centering
   \includegraphics[clip,width=.85\linewidth]{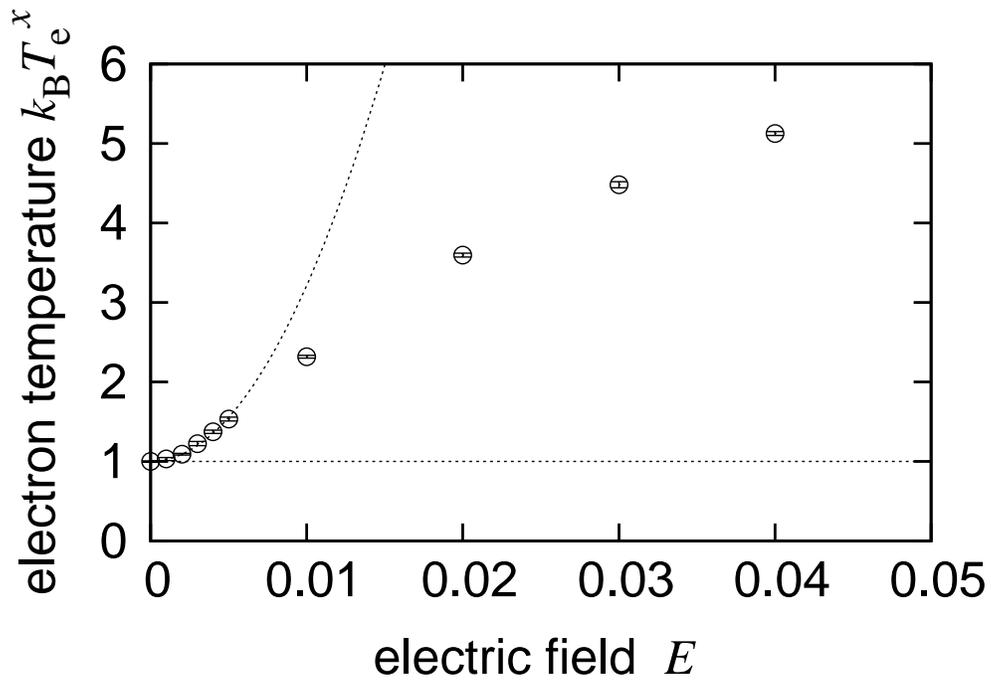}
   \caption{Average values of the electron temperature $k_{\rm{B}} T_{\rm{e}}^x$ against the electric field $E$. 
            The error bars indicate the standard deviation over five impurity configurations. 
            The dotted curve represents eq. (\ref{electron_temperature}) with $b = 2.2 \times 10^4$.
            \label{E_T}}
\end{figure}

\begin{figure}[htbp]
   \centering
   \includegraphics[clip,width=.85\linewidth]{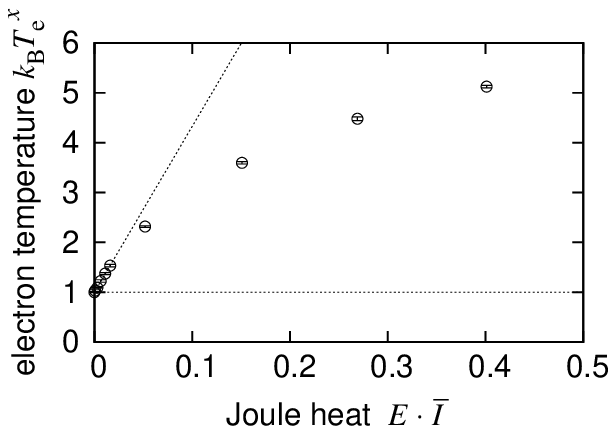}
   \caption{Average values of the electron temperature $k_{\rm{B}} T_{\rm{e}}^x$ against the Joule heat $E \cdot \overline{I}$. 
            The error bars indicate the standard deviation over five impurity configurations. 
            The dotted line represents the linear relation: $k_{\rm{B}} (T_{\rm{e}}^x (E) - T_{\rm{e}}^x (0)) \propto E \cdot \overline{I}$. 
            \label{EI_T}}
\end{figure}

We also investigate the dependence of the electron temperature 
$k_{\rm{B}} T_{\rm{e}}^x \equiv m \langle v_i^x - \langle v_i^x \rangle^2 \rangle$ on the electric field $E$.
As plotted in Fig.~\ref{E_T}, we find that 
\begin{equation}
  k_{\rm{B}} T_{\rm{e}}^x \simeq k_{\rm{B}} T + b E^2 \label{electron_temperature}
\end{equation}
holds for $E \lesssim 0.005$ where $b$ is a positive constant 
(whereas $k_{\rm{B}} T_{\rm{e}}^x$ increases more slowly for $E \gtrsim 0.005$). 
This curve reflects the fact that the Joule heat is proportional to $E^2$ in the linear response regime. 
To see this point, in Fig.~\ref{EI_T}, 
we plot the dependence of $k_{\rm{B}} T_{\rm{e}}^x (E)$ on the Joule heat $E \cdot \overline{I}$. 
For small $E$, it is clearly seen that 
$k_{\rm{B}} (T_{\rm{e}}^x (E) - T_{\rm{e}}^x (0)) \propto E \cdot \overline{I}$, as expected. 
For larger $E$, however, $T_{\rm e}^x$ increases more slowly. 
This means that energy transfer from electrons to phonons becomes very fast for larger $E$. 

\subsection{Dispersion Relations}

We study whether the dispersion relations (Kramers-Kronig relations) 
of the complex conductivity $\sigma(\omega)$ holds in this model.
To obtain $\sigma(\omega)$, we apply an AC electric field $E(t) = E_0 \cos(\omega t)$, 
measure the current during $1000 < t \leq 10000$, 
and calculate the real and imaginary parts of the conductivity 
$\sigma(\omega) = \widetilde{I}(\omega)/ L_y \widetilde{E}(\omega)$, 
where \: $\widetilde{  }$ \: denotes the Fourier transform. 
We set $E_0 = 0.001$, which corresponds to the linear response regime.

For a single impurity configuration, 
we calculate $\sigma(\omega)$ at each $\omega$ 
by averaging the results for 10 samples, 
which have different initial distributions of electrons and atoms, 
as well as different random numbers for the thermal walls. 

\begin{figure}[htbp]
\centering
   \includegraphics[clip,width=.85\linewidth]{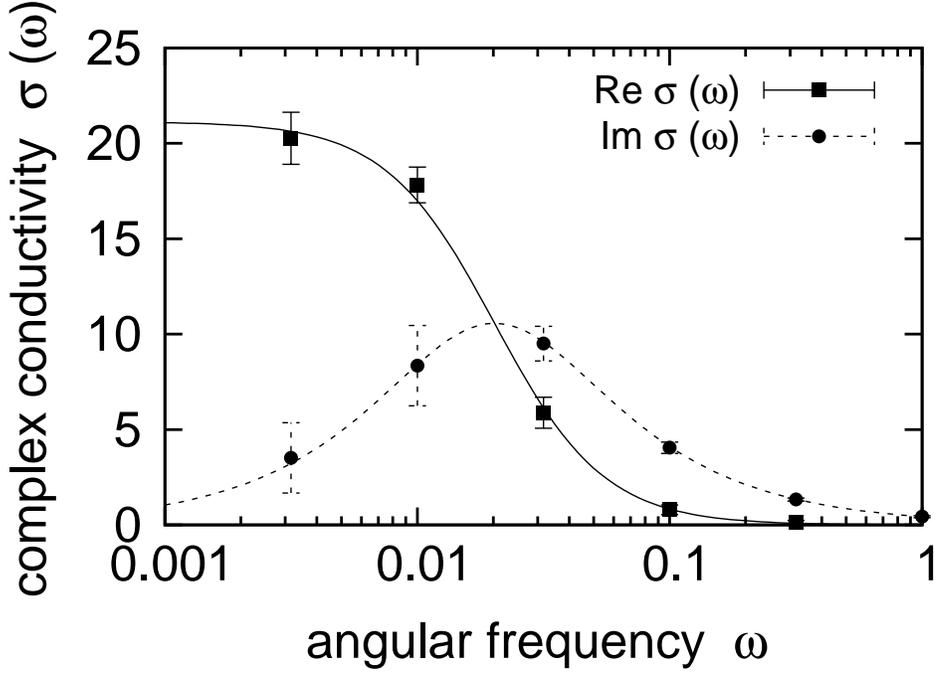}
   \caption{ A semi-logarithmic plot of the real and imaginary parts of the complex conductivity $\sigma(\omega)$ 
             against the angular frequency $\omega$.
             The error bars indicate the standard deviations over 10 samples. 
             The solid and dotted lines are the real and imaginary parts of the fitting curves, eq.~(\ref{Drude_model}). 
             \label{w_sigma}}
\end{figure}

The result is plotted in Fig.~\ref{w_sigma}.
To check the validity of the dispersion relations, 
\begin{align}
 \textrm{Re} \: \sigma(\omega) = 
 \int_{- \infty}^{\infty} \frac{d \omega'}{\pi} \frac{\mathcal{P}}{\omega' - \omega} 
 \: \textrm{Im} \: \sigma(\omega') \: \: , \: \:
 \textrm{Im} \: \sigma(\omega) = 
 - \int_{- \infty}^{\infty} \frac{d \omega'}{\pi} \frac{\mathcal{P}}{\omega' - \omega} 
 \: \textrm{Re} \: \sigma(\omega') \: \: ,
\end{align}
we need to perform the Hilbert transformation. 
However, the transformation is generally difficult 
because it requires knowledge of $\textrm{Re} \; \sigma(\omega)$ and 
$\textrm{Im} \; \sigma(\omega)$ over a very wide range of $\omega$.
Instead, we fit the data to the Drude formula:
\begin{align}
 \textrm{Re} \: \sigma(\omega) = \frac{n_{\rm{e}} e^2 \tau}{m(1 + (\omega \tau)^2)} \: \: , \: \:
 \textrm{Im} \: \sigma(\omega) = \frac{n_{\rm{e}} e^2 \omega \tau^2}{m(1 + (\omega \tau)^2)} \label{Drude_model} \: ,
\end{align}
which satisfy the dispersion relations. 
The fitting parameter is $\tau$.
We see that the data in Fig.~\ref{w_sigma} are well fitted by eq.~(\ref{Drude_model}) with $\tau = 59$.
Therefore, we conclude that the dispersion relation holds.

\subsection{Fluctuation-Dissipation Relation}
\label{sec:FDR}

We also examine whether the fluctuation-dissipation relation (FDR) holds in this model.
Since the complex admittance is given by $\sigma(\omega) L_y / L_x$, 
the FDR is expressed as
\begin{equation}
 S_I(\omega) = 2 k_{\rm{B}} T \; \textrm{Re} \: \sigma(\omega) \cdot \frac{L_y}{L_x}  \: , \label{LHS_RHS}
\end{equation}
where $S_I(\omega)$ is the spectral intensity of the electrical current in the equilibrium state ($E = 0$), 
and $\sigma(\omega)$ is the complex conductivity in the linear response regime ($E_0 = 0.001$). 
We measure the electrical current during $1000 < t \leq 10000$. 
For a single impurity configurations, we obtain the values of the left-hand side (LHS) from 20 samples, 
and the values of the right-hand side (RHS) from 10 samples. 

\begin{figure}[htbp]
\centering
   \includegraphics[clip,width=.85\linewidth]{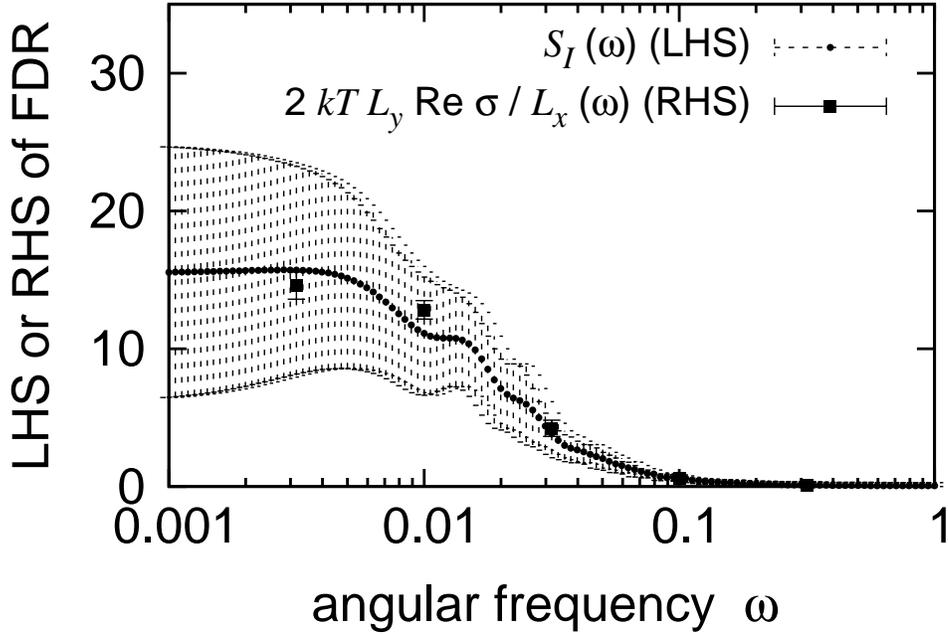}
   \caption{The LHS and the RHS of eq.~(\ref{LHS_RHS}) plotted against the angular frequency $\omega$. 
            The circles represent the LHS and the squares represent the RHS. 
            The error bars indicate the standard deviations over 20 samples for the LHS, 
            and over 10 samples for the RHS. 
   \label{FDR}}
\end{figure}

The results are plotted in Fig.~\ref{FDR}.
We see that the LHS and the RHS agree well at each value of $\omega$.
Therefore, we conclude that the FDR holds in this model.

\subsection{Temperature Dependence of Electrical Conductivity}

We investigate the temperature dependence of the DC electrical conductivity $\sigma(T)$ 
in the linear response regime ($E = 0.001$).

We first examine a system where the electron density is small. 
We take $N_{\rm{e}} = 150$ ($\ll N_{\rm{a}} = 4100$), $N_{\rm{i}} = 0$, $G = 0$, 
$K' = 0$, $G' = 0$. 
In this case many-body effects are expected to be small. 
By measuring the current during $1000 < t \leq 11000$, 
we get $\sigma(T)$. 
We investigate $\sigma(T)$ at $T = 0.125, 0.25, 0.5, 1$, for which the lattice doesn't melt 
according to eq. (\ref{melt}).

\begin{figure}[htbp]
\centering
   \includegraphics[clip,width=.85\linewidth]{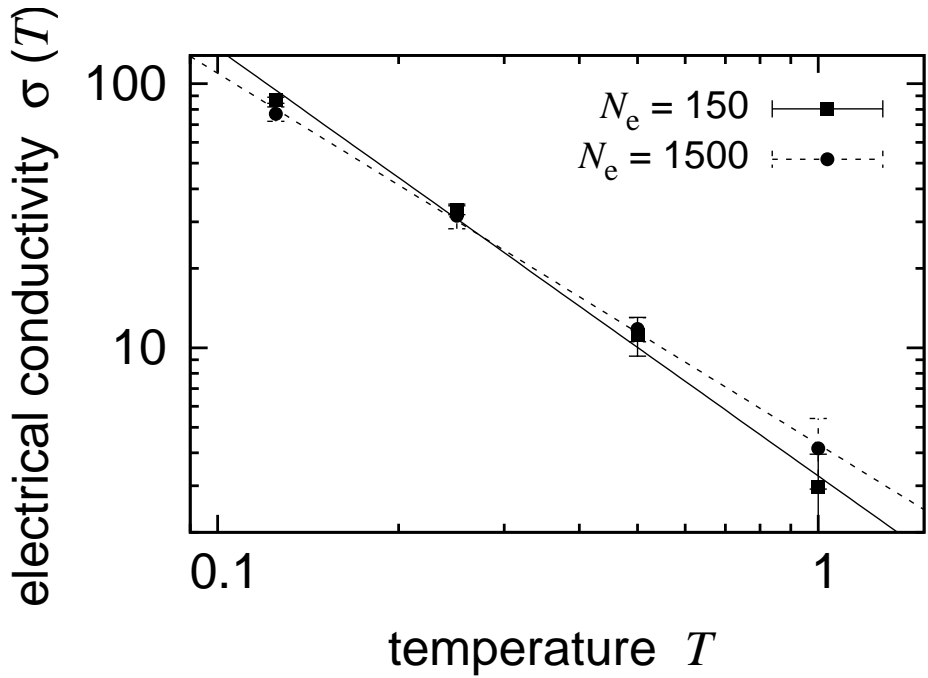}
   \caption{ Temperature dependence of the electrical conductivity for $N_{\rm{e}} = 150$ (squares)
             and for $N_{\rm{e}} = 1500$ (circles). 
             The error bars indicate the standard deviations over five impurity distributions. 
             The solid, and dotted lines are the fitting curves, eq. (\ref{equation_T_sigma}), 
             with $a = 3.27$, $b = -1.61$ for $N_{\rm{e}} = 150$, 
             and $a = 4.31$, $b = -1.41$ for $N_{\rm{e}} = 1500$, respectively. 
             \label{T_sigma}
             }
\end{figure}

The results are shown in Fig.~\ref{T_sigma}. 
The solid line is a fitting curve
\begin{equation}
   \sigma(T) = a \, T^{b} \label{equation_T_sigma}
\end{equation} 
with $a = 3.27 (\pm 0.38)$, $b = -1.62 (\pm 0.09)$. 
The observed exponent $-1.62$ is explained roughly as follows.
If we assume that the long wave length approximation is valid, 
the scattering cross section satisfies 
$s_{\textrm{quantum}}(T) \propto T$.
Since $s_{\rm classical}$ of our model imitates $s_{\rm quantum}$, 
\begin{equation}
\sigma(T) \propto \tau(T) \simeq \frac{l(T)}{\sqrt{\langle v^2 \rangle}_T} 
\simeq \frac{1}{{n_{\rm{a}} \, s_{\textrm{classical}}(T)} \, \sqrt{\langle v^2 \rangle}_T} \: ,
\end{equation} 
holds in the linear response regime, 
where $\tau$ is the mean free time, and $l$ is the mean free path, 
$\sqrt{\langle v^2 \rangle}_T$ is the standard deviation of the speed of the electrons at temperature $T$, 
and $n_{\rm{a}}$ is the density of atoms. 
$n_{\rm a}$ is independent of $T$, 
since the thermal expansion is negligible in this model. 
Therefore, $\sigma(T) \propto T^{-1} T_{\rm{e}}^{-1/2} \, \simeq T^{-3/2}$, 
which is almost consistent with our result. 

We also analyze the case where the electron density is moderate, $N_{\rm{e}} = 1500$ 
($\sim N_{\rm{a}} = 4100$). 
We measure the current during $2000 < t \leq 3000$.
The results are shown also in Fig.~\ref{T_sigma}.
The dotted line is a fitting curve of eq.~(\ref{equation_T_sigma}) 
with $a = 4.31 (\pm 0.20)$, $b = -1.41 (\pm 0.04)$. 
This exponent $b = -1.41 \pm 0.04$ is slightly different from the above case ($b = -1.62 \pm 0.09$).
This difference arises from electron-electron interactions, which are treated on an equal footing with 
electron-phonon interactions in our model.

Those results for the exponent $b$ are consistent with experimental results 
within experimental errors.\cite{Canali:1975, Norton:1973, Morin:1954} 
We have thus obtained reasonable dependence of the electrical conductivity on temperature, 
although our model is a classical model. 
This suggests that we would be able to get the correct temperature dependence of the electrical conductivity 
also in the nonlinear nonequilibrium regime.

%・Debye approximation
%acoustic phononのみだと考えている。
%分散関係をlinearで表して、波数ベクトルの許容範囲をBrillouine zoneと同体積に限定している
%低温で有効（格子系が成り立つ範囲で）
%・Einstein approximation 

%%%%%%%%%%%%%%%%%%%%%%%%%%%%%%%%%%%%%%%%%%%%%%%%%%%%%%%%%%%%%%%%%%%%%%%%%%%%%%%%%%%%%%%%%%%%%

\section{Summary and Conclusion}

In summary, we have proposed a model of electrical conductors, 
which models interacting many electrons scattered by lattice vibrations and random potentials. 
The model is suitable for an MD simulation.

The model consists of classical electrons and atoms. 
The atoms compose a lattice vibration system. 
The electrons are scattered by electron-electron, electron-atom and electron-impurity interactions. 
We have argued that the scattering cross section is physically more important 
than the functional form of a scattering potential. 
Accordingly, we have set the electron-atom interaction potential in such a way that its scattering cross section agrees 
with that of quantum mechanical one. 

To illustrate the advantages of the proposed model, 
we have applied the proposed model to a doped semiconductor at room or higher temperature, 
and have performed an MD simulation. 

In the linear nonequilibrium regime, 
we have confirmed the dispersion relations and the fluctuation-dissipation relation. 
We have also obtained reasonable dependence of the electrical conductivity on temperature, 
despite the fact that our model is a classical model. 

In the nonlinear nonequilibrium regime, NESSs are well realized, 
in which the current $I$ and the kinetic energy $K$ fluctuate 
around the average values with relatively small magnitudes. 
For large $E$, the response of $\overline{I}$ to $E$ becomes strongly nonlinear, 
and the electron temperature raises nonlinearly as a function of the Joule heat. 

Because of these realistic physical properties, 
we expect that the present model should be a good model to explore nonequilibrium states 
of electrical conductors by an MD simulation. 

Although we have studied macroscopically homogeneous conductors in this paper, 
we can also study macroscopically inhomogeneous conductors using the present model. 
For example, we can analyze conductors that are driven by reservoirs, 
where the electrical current is induced by the chemical potential difference of the reservoirs 
attached to the both ends. 
Such systems are suitable for exploring nonequilibrium states, 
because nontrivial and important effects such as nonmechanical forces appear,\cite{Shimizu:1999, Lee:Mthesis}
and may be used to confirm the scaling law of the excess noise. \cite{Footnote2}


\begin{thebibliography}{99}

%%History
\bibitem{Kubo:1995} R.\ Kubo, M.\ Toda and N.\ Hashitsume: \textit{Statistical Physics II}, 2nd ed. 
(Springer, New York, 1995).
\bibitem{Alder:1957}
B.\ J.\ Alder and T.\ E.\ Wainwright: J.\ Chem.\ Phys. {\bf27} (1957) 1208.
\bibitem{Shimada:2000} T.\ Shimada, T.\ Murakami, S.\ Yukawa and N.\ Ito: J.\ Phys.\ Soc.\ Jpn.\ \textbf{69} (2000) 3150.
\bibitem{Yuge:2005} T.\ Yuge, N.\ Ito and A.\ Shimizu: J.\ Phys.\ Soc.\ Jpn.\ \textbf{74} (2005) 1895.
\bibitem{Yuge:2007} T.\ Yuge and A.\ Shimizu: J.\ Phys.\ Soc.\ Jpn.\ \textbf{76} (2007) 093001.
\bibitem{Yuge_alder:2009} T.\ Yuge and A.\ Shimizu: Prog.\ Theor.\ Phys.\ Suppl.\ {\bf 178} (2009) 64.
\bibitem{Yuge:2009} T.\ Yuge and A.\ Shimizu: J.\ Phys.\ Soc.\ Jpn.\ \textbf{78} (2009) 083001.
\bibitem{ShimizuYuge:2010} A.\ Shimizu and T.\ Yuge: J.\ Phys.\ Soc.\ Jpn.\ {\bf 79} (2010) 013002.
\bibitem{Shimizu:2010} A.\ Shimizu: J.\ Phys.\ Soc.\ Jpn.\ {\bf 79} (2010) 113001.
\bibitem{Yuge:2010} T.\ Yuge: Phys.\ Rev.\ E {\bf 82} (2010) 051130.

\bibitem{Shimizu:1999} A.\ Shimizu and H.\ Kato: 
\textit{Low Dimensional Systems - Interactions and Transport Properties}, 
ed. T.\ Brandes (Springer, Berlin, 2000), cond-mat/9911333.

\bibitem{Lee:Mthesis}
F.\ Lee, Master thesis, the University of Tokyo (2009).

%%Model
%\bibitem{Fermi:1955} E.\ Fermi, J.\ R.\ Pasta and S.\ Ulam: Los Alamos Report No.\ LA-1940 (1955), 
%latter published in \textit{E. Fermi, Collected Papers} (Univ.\ Chicago Press, 1965)
\bibitem{Bardeen:1950} J.\ Bardeen and W.\ Shockley: Phys.\ Rev.\ \textbf{80} (1950) 72. 
\bibitem{Ashcroft:1976} N.\ W.\ Ashcroft and N.\ D.\ Mermin: \textit{Solid State Physics}, 
International edition (Holt, Rinehart, and Winston, New York, 1976). 
\bibitem{Shimizu:unpublished} A.\ Shimizu: unpublished.

\bibitem{appendix} One might be tempted to take $U_{\rm ea}$ 
as $U_{\rm ea} = \sum_\alpha \sum_\beta [U (\Vec{r}_{\alpha}, \Vec{R}_{\beta}) - 
                                     U (\Vec{r}_{\alpha}, \Vec{R}_{\beta0})]$,
which imitates effective potential acting on band electrons. 
This potential clearly satisfies the condition that electrons are not scattered 
if atoms are not displaced at lattice points. 
For this form of $U_{\rm ea}$, however, 
we have found that classical electrons are trapped by atoms 
unless the parameters (such as $Q_{\rm{ea}}$ and $l_{\rm{ea}}$) are fine tuned, 
because $U_{\rm ea}$ can take negative values. 
To avoid such unphysical effects, we did not adopt this form of $U_{\rm{ea}}$ in this paper.

%%Semiconductor
\bibitem{Wolfe:1970} C.\ M.\ Wolfe, G.\ E.\ Stillman, and W.\ T.\ Lindley: J.\ Appl.\ Phys.\ \textbf{41} (1970) 3088. 
\bibitem{Ando:1982} T.\ Ando, A.\ Fowler and F.\ Stern: Rev.\ Mod.\ Phys.\ \textbf{54} (1982) 437.

%%MD
\bibitem{Quentrec:1975} B.\ Quentrec and C.\ Brot: J.\ Comput.\ Phys.\ \textbf{13} (1975) 430. 
\bibitem{Hatano:1998} T.\ Hatano: Phys.\ Rev.\ E.\ \textbf{59} (1999) R1. 
\bibitem{Tehver:1998} R.\ Tehver, F.\ Toigo, J.\ Koplik and J.\ R.\ Banavar: Phys.\ Rev.\ E.\ \textbf{57} (1998) R17.

%%Results1.
\bibitem{Canali:1975} C.\ Canali, C.\ Jacoboni, F.\ Nava, G.\ Ottaviani, and A.\ Alberigi-Quaranta: Phys.\ Rev.\ B.\ \textbf{12} (1975) 2265.
\bibitem{Norton:1973} P.\ Norton, T.\ Braggins, and H.\ Levinstein: 
Phys.\ Rev.\ B.\ \textbf{8} (1973) 5632.
\bibitem{Morin:1954} F.\ J.\ Morin and J.\ P.\ Maita: 
Phys.\ Rev.\ \textbf{96} (1954) 28.

%%Summary
\bibitem{Footnote2} In such a case, it is better to take account of the long-range Coulomb potential, 
because electrons are accumulated or depleted in the regions close to the reservoirs.

\end{thebibliography}
\end{document}